\documentclass[twocolumn]{aastex631}

\newcommand{\snrf}{F$_{\mathrm{S/N}}$}
\newcommand{\snre}{E$_{\mathrm{S/N}}$}
\newcommand{\fitf}{F$_{\mathrm{fit}}$}
\newcommand{\fite}{E$_{\mathrm{fit}}$}
\shorttitle{Effects of banded repeaters}
\shortauthors{Aggarwal}
\graphicspath{{./}{figures/}}

\begin{document}

\title{Observational effects of banded repeating FRBs}

\correspondingauthor{Kshitij Aggarwal}
\email{ka0064@mix.wvu.edu}

\author[0000-0002-2059-0525]{Kshitij Aggarwal}
\affil{West Virginia University, Department of Physics and Astronomy, P. O. Box 6315, Morgantown, WV, USA}
\affil{Center for Gravitational Waves and Cosmology, West Virginia University, Chestnut Ridge Research Building, Morgantown, WV, USA}



\begin{abstract}
Recent observations have shown that repeating Fast Radio Bursts (FRBs) exhibit band-limited emission, whose frequency-dependent amplitude can be modeled using a Gaussian function. In this analysis, we show that banded emission of FRBs can lead to incompleteness across the observing band. This biases the detected sample of bursts and can explain the various shapes of cumulative energy distributions seen for repeating FRBs. We assume a Gaussian shape of the burst spectra and used simulations to demonstrate the above bias using an FRB~121102-like example. We recovered energy distributions that showed a break in power-law and flattening of power-law at low energies, based on the fluence threshold of the observations. We provide recommendations for single-pulse searches and analysis of repeating FRBs to account for this incompleteness. Primarily, we recommend that burst spectra should be modeled to estimate the intrinsic fluence and bandwidth of the burst robustly. Also, bursts that lie mainly within the observing band should be used for analyses of energy distributions. We show that the bimodality reported in the distribution of energies of FRB~121102 by \citet[][]{di_121102} disappears when burst bandwidth, instead of the center frequency of the observation, is used to estimate energy. Sub-banded searches will also aid in detecting band-limited bursts.  All the analysis scripts used in this work are available in a Github repository\footnote{\url{https://github.com/KshitijAggarwal/banded_repeater_analysis}}. 
\end{abstract}

\keywords{Radio transient sources(2008) --- Extragalactic radio sources(508) --- Radio bursts(1339)}



\section{Introduction} \label{sec:intro}

Fast radio bursts (FRBs) show a wide variety of properties \citep{frb_review}. It is an ongoing effort to disentangle the properties arising from the FRB source itself (i.e., intrinsic) to those introduced due to various selection effects. FRBs that emit multiple bursts, called repeaters, appear to be distinct in their observational characteristics as compared to the apparent non-repeaters \citep{chime_cat, chime_morph}. Two properties associated primarily with repeating FRBs are sub-burst drifting and band-limited emission \citep[][]{hessels2019, Law2017}. 
In this Letter, the effect we focus on is the band-limited emission of repeaters and discuss various observational biases caused by it. 

The source of the band-limited nature of FRB emission is currently not understood. However, there exist multiple progenitor and propagation models that try to explain the band-limited nature of FRBs \citep{Simard2020, paz2020, Metzger2019, lu2018, cordes2017}. Further, it has been reported that the emission of the first repeating FRB, FRB~121102 favors 1600 MHz, and there is a lack of emission observed below 1200 MHz \citep{k_121102, platts2021}. It is unclear if the emission behavior changes below that frequency and is below the detection threshold or is just not present \citep{platts2021}. Similar banded\footnote{We use ``band-limited" and ``banded" emission interchangeably across this Letter. Both of these refer to the finite bandwidth emission of the bursts.} emission has also been reported for other repeating FRBs \citep{Marazuela2020, di_121102, k_121102}. Further, it has been reported that the peak emission frequency appears to be random. Also, ultra-wideband observations have shown that there is no evidence for oscillations in the spectra, i.e the burst emission is present only within the narrow envelop and is not simultaneously present at any other frequency  \citep[][]{Kumar2021, Law2017}. The periodicity in burst activity and its frequency dependence further complicates the interpretation of this effect \citep[][]{Marazuela2020, rfr3}.

Energy distribution of repeating FRBs also show a variety of shapes and features like: a simple power-law, broken power-law, smooth flattening of power-law at low energies, etc \citep[][]{Cruces2020, di_121102, k_121102, Marazuela2020}. The shape of the energy distribution can provide important information about the emission mechanism of the FRB source. Giant pulses from neutron stars and high energy emission from magnetars has been described using a power-law distribution \citep[][]{Bera2019, Cheng2020}. A similar power-law index for FRBs with that seen from neutron stars might imply a common origin. In some studies of repeaters, the deviation of energy distribution from a simple power-law has been attributed to the intrinsic emission process of the FRB. But, in this Letter, we demonstrate that many of the observed shapes of the energy distributions, can be attributed to biases due to the band-limited nature of emission of repeaters. It is therefore necessary to account for these biases, before making conclusions about the intrinsic nature of FRB emission.



We start by discussing our modeling methods in \S2, followed by results demonstrated using an example in \S3, and discuss the implications in \S4. We then conclude with some recommendations in \S5.    


\section{Methods}
This section briefly discusses the methods we used to model the FRB spectra and various observational effects. We follow a two-step approach: 1) Simulate a population of FRBs with spectral properties described by predefined functions, 2) Estimate the detectable FRBs from this sample based on a sensitivity threshold. 

\subsection{Generating a population of repeater bursts}
\label{sec:genfrbs}
Previous studies have shown that the spectra of repeating FRBs can be modeled using a Gaussian function \citep[][]{Pleunis2021, k_121102, Kumar2021, Law2017}. Therefore, we assume that the repeater burst's spectra follow a Gaussian function, parameterized by a mean ($\mu_f$) and a standard deviation ($\sigma_f$). We also assume that $\mu_f$ and $\sigma_f$ itself follow a predefined distribution (Gaussian or Uniform). Finally, we assume that the cumulative energy distribution follows a power-law with a slope ($\alpha$). We then draw 50,000 samples from the above three distributions each to represent 50,000 bursts from a repeater. We convert the energy into an intrinsic fluence using its spectral bandwidth and assuming a nominal distance to the source. 

\subsection{Applying selection effects}
\label{sec:selection_effects}

\begin{figure*}
    \centering
    \includegraphics[width=\textwidth]{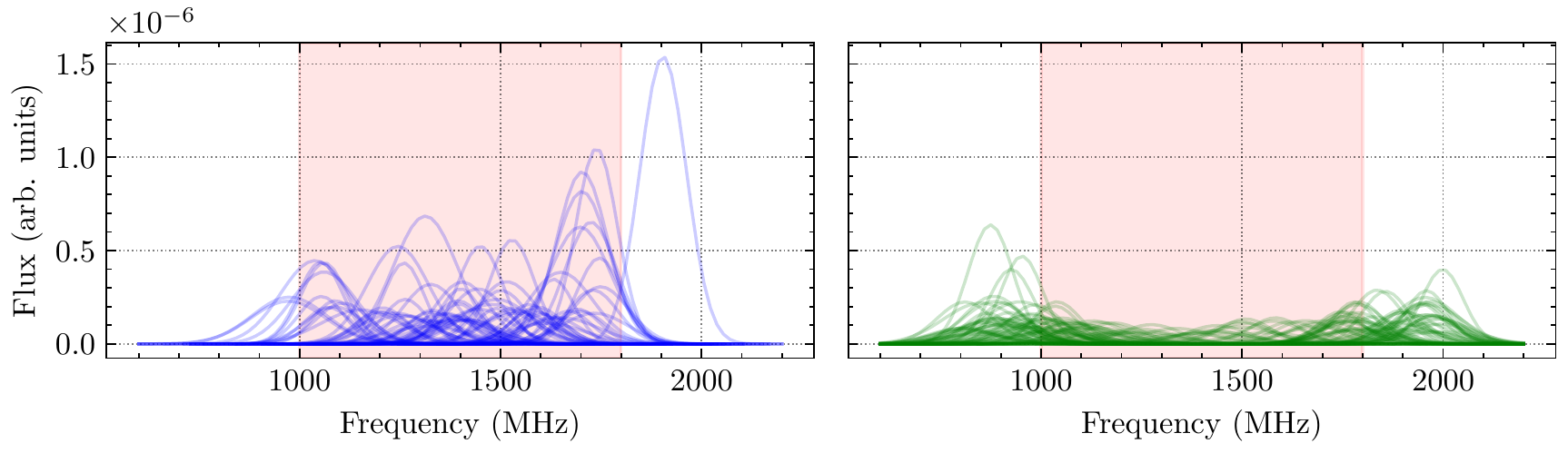}
    \caption{Figure showing some simulated spectra modeled using a Gaussian function. The mean and standard deviation has been sampled from a Uniform distribution. The energy distribution is assumed to be a power-law with a slope of -1.8. The left panel shows the detected spectra, while the right panel shows the ones that weren't detected, using a constant fluence threshold. The observing band is shown in red. Spectra that had enough energy within the band were detected, while the ones without enough signal within the observing band were not detected. See Section~\ref{sec:selection_effects} for more details.}
    \label{fig:spectra_vis}
\end{figure*}

Next, we aim to determine the bursts that a given observational system will detect. To simplify the analysis, we ignore the effect of any signal lost due to non-idealized dispersion measure (DM) and boxcar search step \citep[see][on details of those effects]{k_121102, rfclustering, gb, keane2015}, and only consider two selection effects: observational bandwidth and fluence limit. Due to the limited observing bandwidth of most instruments and the distribution of burst spectra across a wide frequency range (see Figure~\ref{fig:spectra_vis}), the observed fluence of the burst will depend on the fraction of the burst spectra that lies within the observational band. Therefore, for each simulated burst, we estimate its observed fluence by integrating the spectra (using the bursts'  $\mu_f$ and $\sigma_f$) within the limits of observing bandwidth. Therefore, assuming Gaussian spectra of the form, 
\begin{eqnarray}
    \mathcal{G}(\nu;\mu_\nu, \sigma_\nu) = \frac{1}{\sigma_\nu \sqrt{2 \pi}} \exp{\bigg(-\frac{1}{2} \frac{(\nu-\mu_\nu)^2}{\sigma_\nu^2}\bigg)}
\end{eqnarray}

where $\nu$ is the observing frequency. The observed fluence is given by: 

\begin{eqnarray}
    S_{obs} = S_{int} \int_{\nu_{start}}^{\nu_{end}} \mathcal{G}(\nu;\mu_\nu, \sigma_\nu) d\nu
\end{eqnarray}

Where $S_{int}$ and $S_{obs}$ are the intrinsic and observed burst fluence, $\nu_{start}$ and $\nu_{end}$ are the start and end frequency of the observing band. If the burst spectra lie primarily within the bandwidth, then the integral in the above equation will be close to one, and so the observed fluence will be very similar to the intrinsic fluence. As the fraction of burst signal outside the band increases, the observed fluence will get lesser as compared to intrinsic fluence. 

If this observed fluence is greater than the fluence threshold of the search pipeline, then the burst will be detected. 
Therefore, out of the population of bursts that do not lie primarily within our observing band, we are sensitive to detect only the bright ones (as illustrated in Figure~\ref{fig:spectra_vis}). This will introduce strong selection effects in the properties of the detected bursts. 


\subsection{Estimating Fluence and bandwidth}
\label{sec:fluence}
Typically, the detected bursts' fluence (and then energy) is determined only from the signal visible in the observing band. This is done using the signal-to-noise (S/N) ratio obtained from the de-dispersed and frequency averaged time-series profile of the burst. However, this procedure underestimates the intrinsic fluence of the burst, as it is sensitive to signal present only in the observing band. On the other hand, if we model the observed spectra using a Gaussian function, then the total fluence (including the signal not visible in the band) can be estimated and determined \citep[see section 3.5 of][]{k_121102}. Using this burst modeling, it is possible to obtain a more robust estimate of the intrinsic fluence of the burst (i.e., $S_{int}$). 

Similarly, the bandwidth of the bursts is typically determined by manually identifying the range of frequency channels in which the burst signal is visible. For band-limited bursts that lie on the edge of the band, this would lead to underestimation of the burst bandwidths. Again, suppose we model the burst spectra using a Gaussian model. In that case, we can accurately determine the full width at half max (FWHM) of all the bursts robustly, even if only a small fraction of burst spectra is within the observing band.

We use the following equation to estimate the isotropic energy of the bursts from the fluence (S), bandwidth (FWHM), and distance (D$_\mathrm{L}$) of the bursts

\begin{eqnarray}
\label{eq:energy}
    {E} = && 4\pi 10^{-23}
    \left(\frac{\mathrm{D}_\mathrm{L}}{\mathrm{cm}}\right)^{2}
    \left(\frac{\mathrm{S}}{\mathrm{Jy~s}}\right)
    \left(\frac{\mathrm{FWHM}}{\mathrm{Hz}}\right) \mathrm{erg}.
\end{eqnarray}

Henceforth, we refer to the S/N derived fluence as \snrf, and energies estimated using \snrf\ and manually identified bandwidth as \snre. We will use \fitf\ and \fite\ when the fluence and energy are determined using fitting. 

The above three steps give us a sample of simulated and detected repeater bursts that can now be compared to infer the observational biases and incompleteness, with some assumptions on the intrinsic ($\mu_f$, $\sigma_f$, $\alpha$) and observational (fluence threshold and bandwidth) parameters. This is explored in the next section.

\section{Results}
\label{sec:results}
Here we consider a simple example to report some of the observed effects of the band-limited nature of burst spectra. We consider an FRB~121102-like repeater observed at varying detection thresholds. We assume the following intrinsic properties for this repeater: cumulative energy distribution follows a single power-law with a fixed slope ($\alpha$=-1.5), a distance of 972~Mpc, normal distribution of $\mu_f$ with mean 1650~MHz and standard deviation of 250~MHz, and normal distribution of $\sigma_f$ with mean of 300~MHz and standard deviation of 250~MHz. We also assume that our observing bandwidth is 800~MHz with a center frequency of 1375~MHz. The choice of these values is inspired by the observed properties of FRB~121102 reported by \citet{k_121102}. We then observe this repeater at three different fluence limits (i.e. sensitivity limit of the observing system) of: 0.02~Jy\,ms, 0.1~Jy\,ms and 0.4~Jy\,ms. 
As discussed previously, some of these bursts will not be detected by the search system, as not enough burst signal is present within the band. In the following sub-sections, we discuss three unique observational effects observed for this simulated repeater. We also try to explain the various observed properties of the two most studied repeaters so far: FRB~121102 and FRB~180916, in the context of the incompleteness due to the banded nature of their bursts. 


\subsection{Cumulative Energy Distribution}
\label{sec:cumulative_e}
Many different shapes and slopes have been reported for the cumulative energy distribution of repeater burst energies \citep{k_121102, Marazuela2020, Cruces2020, Law2017}. 
Here we discuss how telescope sensitivity plays a crucial role in determining the observed shape of the cumulative energy distribution by influencing the bursts that are detected by the system. 

\begin{figure*}
    \centering
    \includegraphics[width=\textwidth]{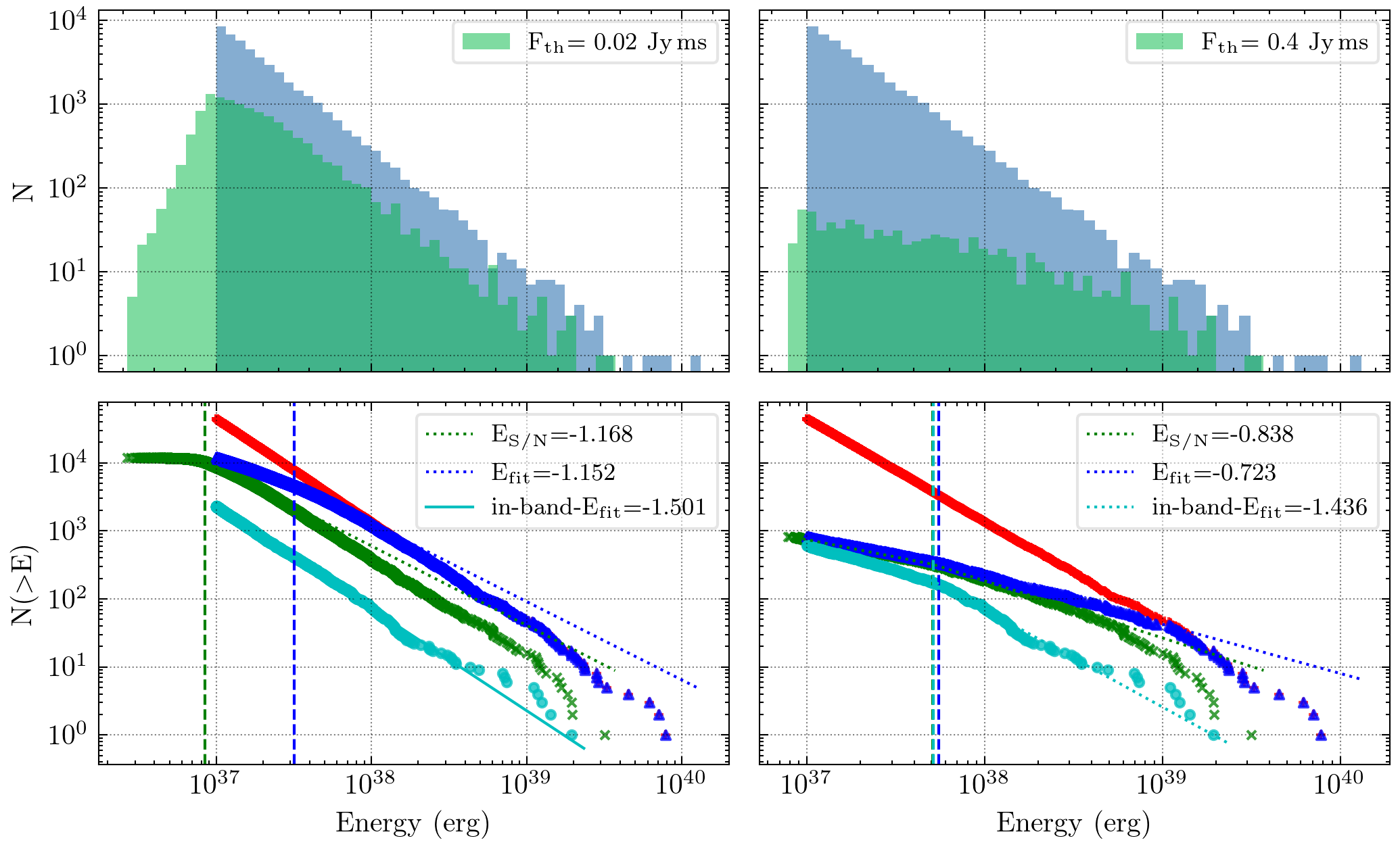}
    \caption{Energy distributions of bursts detected at varying fluence thresholds. Different columns show different fluence thresholds - left: 0.02~Jy\,ms, right: 0.4~Jy\,ms. Top: Histogram of burst energies. Intrinsic energies of the bursts are shown in blue and show the single power-law, while those of detected bursts are shown in green. In this case, the burst energies are estimated using \snrf\ (i.e using fluence derived from signal to noise of the burst; see Sections~\ref{sec:cumulative_e} and \ref{sec:energy_pdf}). Bottom: Cumulative energy distribution of bursts. Red pluses show the intrinsic energy distribution that follows a power-law with a slope of -1.5. The other three colors show detected bursts. Green crosses represents energies estimated using \snrf. Blue triangles and cyan circles are with energies estimated using fitting. Blue triangles shows all the detected bursts, while cyan circles only shows bursts that were primarily within the observing band. Solid lines show a single power-law fit, and dotted lines show a broken power-law fit. The vertical dashed lines shows the break energy for the broken power-law fit. Values in legends report the fitted slope for single power-law fit, and high energy slope for broken power-law fit. As we can see, due to inaccurate energy estimation and missed bursts, the energy distributions can deviate significantly from the intrinsic distribution. See Section~\ref{sec:energy_pdf} for more details.}
    \label{fig:energy_pdf}
\end{figure*}

The bottom row of Figure~\ref{fig:energy_pdf} shows the cumulative energy distribution of bursts detected with two different fluence thresholds. Red pluses show the intrinsic power-law, green crosses show the \snre\ for the detected bursts, and blue triangles represent the \fite\ of the detected bursts. The trend of blue triangles and green crosses points deviate significantly from a simple intrinsic power law due to the absence of weak bursts that are not detected. Notably, the shape of observed cumulative energy distribution changes with sensitivity thresholds. As the sensitivity of the observations decrease (left to right in Figure~\ref{fig:energy_pdf}), the observed distributions deviate from a single power-law to a broken power-law, which appears as a smooth turn at even lower thresholds. The broken power-law is similar to what has been seen for FRB~121102 \citep{Cruces2020, k_121102} while the smooth turn in power-law was reported for FRB~180916 \citep{Marazuela2020}. Here, we have shown that both these effects can occur due to the band-limited nature of repeater bursts.

\subsubsection{Challenges}
Various possible shapes of the observed cumulative distribution make it challenging to interpret and infer intrinsic FRB properties. A break in the power-law, if present, might indicate the real completeness limit of the system. This can be observed by the green crosses in the bottom-left plot of Figure~\ref{fig:energy_pdf}. The higher energy slope appears similar to the intrinsic one.  Therefore, a break in observed power-law might indicate that the higher energy power-law follows the intrinsic shape and may be used to draw inferences about the intrinsic properties of the FRB. A smooth turnover in cumulative distribution (right panel of Figure~\ref{fig:energy_pdf}), on the other hand, renders a completeness limit derived from such an analysis inaccurate, and even the higher energy slope might not represent the intrinsic distribution of energies. Even the distribution of \fite\ might not reflect the real power-law, as the detection threshold would also bias it. 

\subsubsection{Using in-band bursts}
When using \fite, although we are using the correct fluences of the bursts, we still would have missed a population of weak bursts that were not primarily within our observing band (see Section~\ref{sec:selection_effects}). Therefore, we would detect a larger sample of bright bursts, which will make the energy distribution flatter. We can account for this by analyzing only the in-band bursts, i.e., the bursts whose spectra lie primarily within our observing band
\footnote{A similar condition was also used by \citet{k_121102} to select the bursts for cumulative energy distribution analysis.}. These bursts are labeled as in-band-\fite and shown as cyan circles in Figure~\ref{fig:energy_pdf}. These in-band bursts provide a more robust and reliable estimate of the distribution of burst energies, even when the observations are not very sensitive. This is because our observations are complete to the in-band bursts, and by using fitting to determine burst properties, we have mitigated both selection effects listed earlier (see Section~\ref{sec:selection_effects}). The slope of the in-band burst energies follows the intrinsic distribution. But, if the distribution shows a turnover or flattening, then even this method cannot be used to estimate the intrinsic power-law slope of the bursts reliably.

\subsection{Energy Distribution}
\label{sec:energy_pdf}
Continuing the previous example, the top row in Figure~\ref{fig:energy_pdf} shows the histogram of \snre\ for two fluence thresholds. Again, the intrinsic cumulative distribution of burst energies follows a power-law. The detected energy distribution again changes significantly with the fluence threshold, and even in this simple case, it shows a variety of shapes. 
All of these shapes can be attributed to the missed bursts and inaccurate estimates of burst energies. Using \fite\ leads to distributions that are more similar to the intrinsic one, and using \fite\ for only in-band bursts would be an even more accurate representation of intrinsic distribution. 

\subsection{Distribution of spectral parameters}
\label{sec:spectra_dist}

\begin{figure*}
    \centering
    \includegraphics[width=\textwidth]{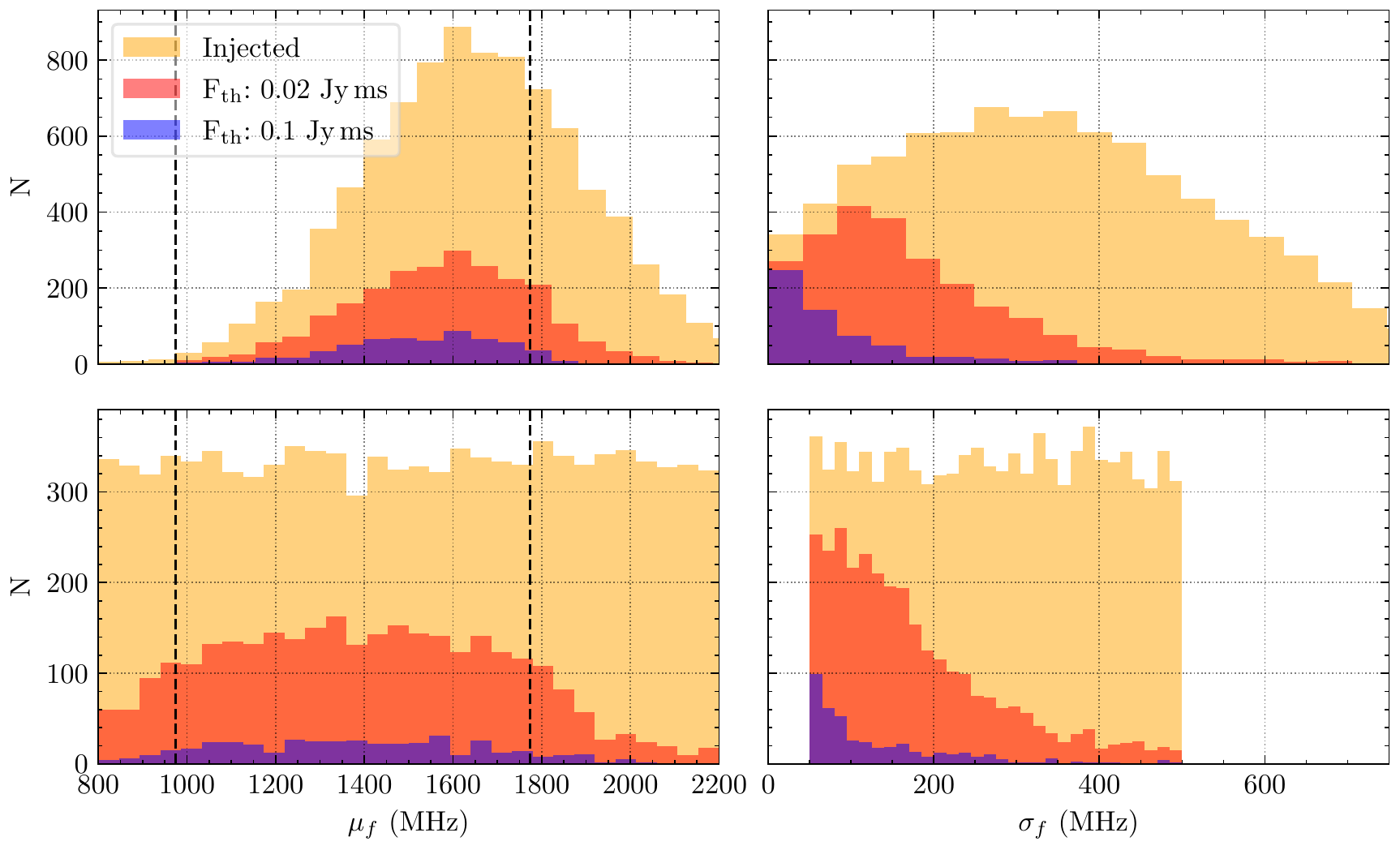}
    \caption{Distribution of mean ($\mu_f$) and standard deviation ($\sigma_f$) of burst spectra. The intrinsic distribution is shown in yellow, and the distribution of bursts detected at various fluence thresholds is shown in red and blue. The top panels consider a normal intrinsic distribution of $\mu_f$ and $\sigma_f$, while bottom panels assume a uniform intrinsic distribution. Vertical dotted lines mark the observing band. See Section~\ref{sec:spectra_dist} for more details.}
    \label{fig:spectra_param}
\end{figure*}

We consider two choices of intrinsic distributions for both $\mu_f$ and $\sigma_f$: Uniform and Normal, and observed the respective distributions of the detected bursts (see Figure~\ref{fig:spectra_param}). The $\mu_f$ distribution of detected bursts follows a normal-like distribution for both intrinsic distributions. $\mu_f$ distribution for FRB~121102 has been reported to be an asymmetric normal distribution, with a negative skew, i.e., with a tail towards lower frequencies being drawn out \citep{k_121102}. The FRB emission also prefers higher frequencies in the 1.4~GHz observations.
This cannot be recovered using a uniform intrinsic distribution of $\mu_f$, as in this case, the recovered distribution would peak at the center of the observing band (bottom left panel in Figure~\ref{fig:spectra_param}). On the other hand, if the intrinsic $\mu_f$ distribution is normal, with its mean present towards the top of the center frequency of the observing band, then an asymmetric normal distribution is recovered. The recovered $\mu_f$ distribution of our simple example also shows a negatively skewed distribution (top left panel in Figure~\ref{fig:spectra_param}). This is because fewer bursts will be detected towards the higher part of the band. In this case, the peak of the recovered distribution will lie close to the peak of the intrinsic distribution. Hence, based on results presented in \citet{k_121102}, we can infer that the intrinsic $\mu_f$ distribution of FRB~121102 bursts could be normal, with a mean of $\sim1600$~MHz. 

The observed normal distribution of $\sigma_f$ reported by \citet{k_121102} can also be explained only by an intrinsic normal distribution of $\sigma_f$ and is not recovered by a uniform distribution of $\sigma_f$. This can also be seen in the right panels in Figure~\ref{fig:spectra_param}.

\subsection{Calculating Energy}
\label{sec:calc_energy}
As mentioned previously, we used Eq.~\ref{eq:energy} to estimate the energy of the burst. This equation uses the burst bandwidth along with the fluence to estimate the energy. This formalism is the way to estimate the burst energy when the emission is not broad-band. Under the assumption that emission is broad-band, a standard technique to estimate energy from fluence uses center frequency of the band, instead of burst bandwidth \citep[][]{zhang2018}. This is given by, 

\begin{eqnarray}
\label{eq:energy_centerfreq}
    {E} = && 4\pi 10^{-23}
    \left(\frac{\mathrm{D}_\mathrm{L}}{\mathrm{cm}}\right)^{2}
    \left(\frac{\mathrm{S}}{\mathrm{Jy~s}}\right)
    \left(\frac{\nu_c}{\mathrm{Hz}}\right) \mathrm{erg}.
\end{eqnarray}

where $\nu_c$ is the center frequency of the observing band. Recently, \citet[][]{di_121102} used this latter definition of energy and found that the burst energy of FRB~121102 follows a bimodal distribution, using a large sample of bursts. We re-calculated the burst energies using the burst bandwidths and fluences reported in their Supplementary Table 1 \citep[][]{di_121102} and compared them with the energies used by \citet[][]{di_121102}. 

\begin{figure}
    \centering
    \includegraphics[width=0.48\textwidth]{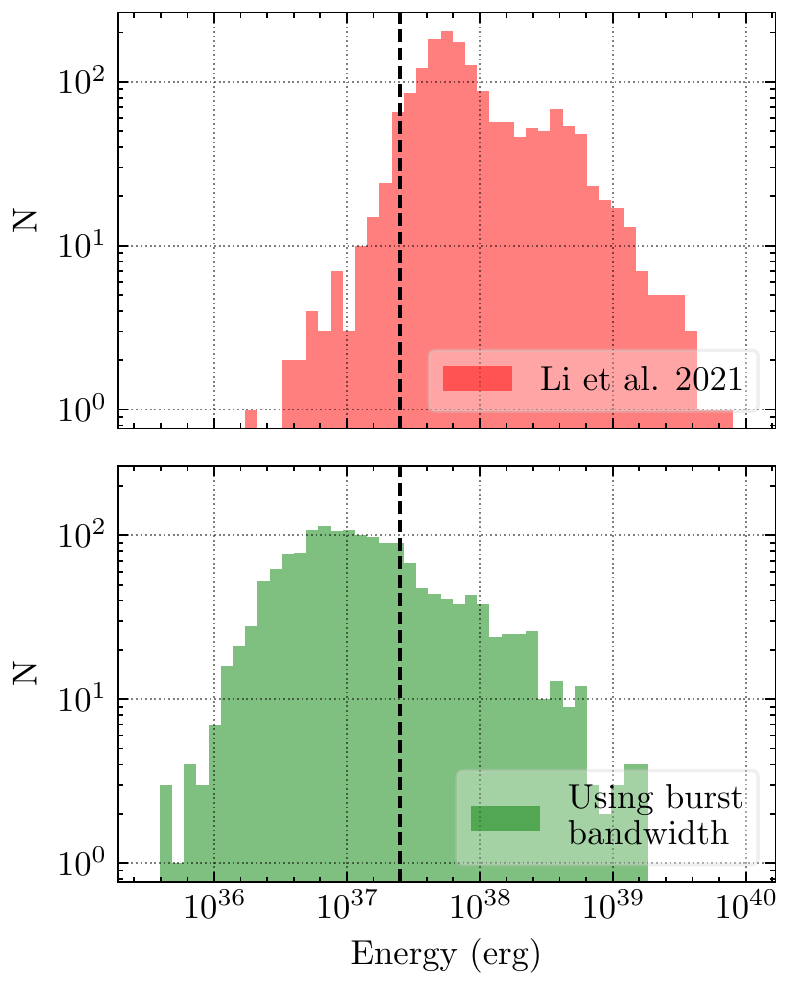}
    \caption{Energy distribution of FRB~121102 bursts reported by \citet[][]{di_121102}. The top panel shows energies calculated using the center frequency of the band (i.e., 1.25~GHz), while the bottom panel shows energies calculated using the bandwidths of the bursts. Vertical dashed line represents the 90\% completeness limit of FAST observations estimated by \citet[][]{di_121102}. The bottom distribution does not show the bimodality seen in the top figure.}
    \label{fig:fast_121102}
\end{figure}

Figure~\ref{fig:fast_121102} shows the distribution of energies calculated using these two techniques. The distribution of energies does not show any bimodality when burst bandwidths, instead of center frequency, are used to estimate energies. Moreover, the resulting distribution of energies is similar to the ones shown in Figure~\ref{fig:energy_pdf}, implying that this result is likely affected by band incompleteness. 
The method of using center frequency to estimate burst energy makes two key assumptions: 1) Emission is broadband, 2) Spectral index is zero, i.e the emission does not depend on frequency. But, the emission from repeaters is characteristically band-limited and Gaussian-like. Therefore, none of these two assumptions are valid for repeating FRBs and so burst bandwidths should be used to estimate energy.

\begin{figure*}
    \centering
    \includegraphics[width=\textwidth]{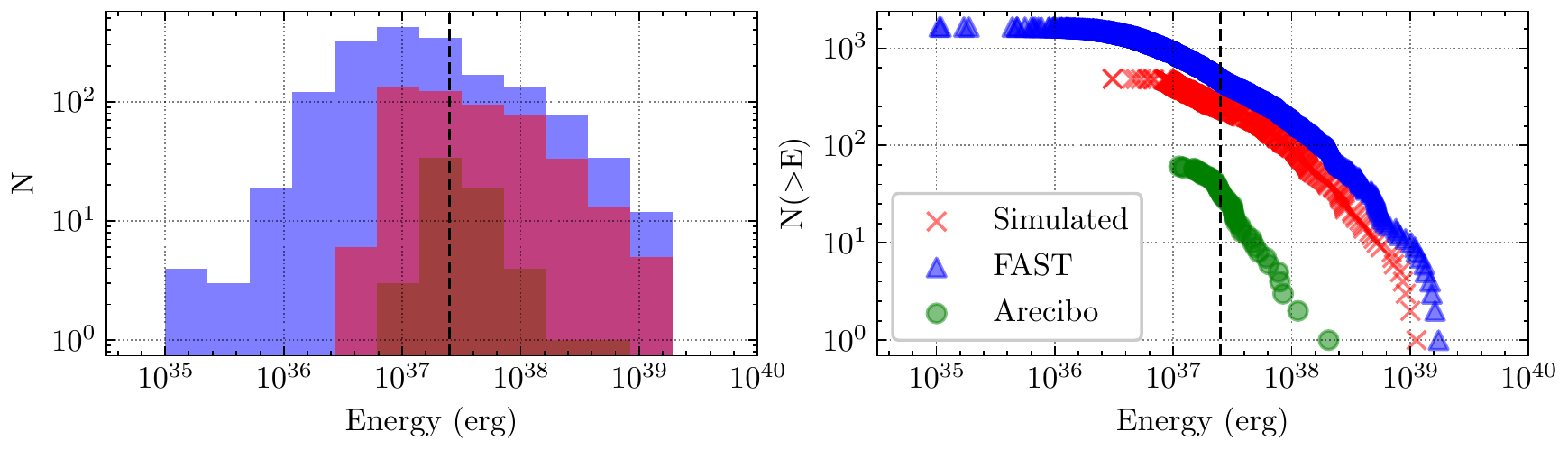}
    \caption{Energy distributions of repeater bursts. The left panel shows the histogram of energies, and the right panel shows the cumulative distribution of those energies. Blue triangles represent FRB~121102 bursts reported by \citet[][]{di_121102}, where we used burst bandwidth to estimate the burst energy. Green circles show the in-band FRB~121102 bursts reported by \citet[][]{k_121102}. Red crosses show the bursts that were detected using the simulation example discussed in Section~\ref{sec:results}. Black dashed lines shows the 90\% completeness limit of FAST observations \citep[][]{di_121102}. The distribution of the observed FRB~121102 energies is very similar to those recovered using simulations. See Section~\ref{sec:calc_energy} for more details.}
    \label{fig:121102_energies}
\end{figure*}

Figure~\ref{fig:121102_energies} shows the distribution of energies (derived using burst bandwidths) of FRB~121102 bursts detected with FAST and Arecibo \citep[][]{k_121102, di_121102}, along with the recovered distribution of detected bursts using some fiducial values for intrinsic properties similar to the simulation example discussed earlier. Therefore, our analysis shows that the observed shapes in the energy distributions of FRB~121102 can occur due to observational effects caused by bandedness along with a normal distribution of the peak and width of the burst spectra. Notably, both these effects were reported for the bursts presented in \citet[][]{di_121102, k_121102}.

We performed power-law fits on the cumulative distribution of energies (derived using burst bandwidths) from FRB\,121102 bursts detected by FAST. We only used the bursts above the energy of $1.2\times10^{37}$\,ergs, estimated from the 95\% completeness limit (0.06~Jy\,ms) of FAST observations reported by \citet[][]{di_121102} and mean burst bandwidth of FAST bursts (200~MHz). We fitted these bursts using a single power-law and a broken power-law. The fitted slope for single power-law fit was $-0.716\pm0.002$. The slopes (below and above the break energy) for the broken power-law fit were $-0.693\pm0.001$ and $-1.12\pm0.02$ with a fitted break energy of $(1.05\pm0.02) \times10^{38}$\,ergs. The high energy slope is consistent with the results of \citet[][]{Cruces2020}, while they are inconsistent to those reported by \citet[][]{k_121102}. It should be noted that the energies of FAST bursts were not corrected for the incompleteness due to banded nature of bursts. Further, the fluences were derived using S/N and not using fitting. As discussed, both these effects could lead to the incorrect estimation of energy distribution for this sample.

\section{Discussion}
The results presented in the previous section caution against estimating completeness limits without accounting for the variety of spectral properties of the FRB, especially repeaters. It further complicates the interpretation of energy distributions and properties intrinsic to the FRB. Two main challenges stand out due to the band-limited nature of repeater bursts: 1) Detecting the band-limited bursts, 2) Robust estimation of fluence and bandwidth of bursts that are only partially within the observing band. In the following subsections, we discuss these two challenges. 


\subsection{Observing bandwidth and sub banded searches}
One possible technique to mitigate the first challenge listed above is to have a large observing bandwidth. A larger observing bandwidth would make it more likely for the burst spectra to fall within the observing band and aid in detecting more bursts.  

Another way is to perform subband searches instead of searching full observational bandwidth \citep[R. Anna-Thomas et al. 2021, in prep; ][]{Kumar2021}. The observing band is divided into multiple subbands, on which a single-pulse search is then performed. This strategy is more sensitive to weak band-limited single pulses that will not exceed the detection threshold with a more traditional full-band single-pulse search. 

For the example presented in the previous section, we estimated the increase in the number of detected bursts using a subband search as compared to a full band search. We divided the observing band into four subbands of 200~MHz each and then performed the single-pulse search. We detected more bursts for all the three fluence threshold cases (an increase of 5, 16, and 41\%). More bursts, as expected, reduced the observational biases and resulted in recovered properties that were more representative of the intrinsic distributions.

\subsection{Calculating fluences and bandwidths}
The second challenge listed earlier was a robust estimation of fluence and bandwidth of detected bursts. 
There is no agreed-upon method to estimate the fluence of the bursts as mentioned previously (see Section~\ref{sec:selection_effects}); signal within the observing band is typically used to estimate burst fluence. In some cases, this signal has been modeled using a power-law, a Gaussian, or a running power-law \citep[][]{chime_cat, Pleunis2021, k_121102} to estimate the burst fluence and other properties. This is because different FRBs show different spectral properties. In this analysis, we have assumed repeater spectra to be Gaussian, as has been reported by observational campaigns on the most studied repeaters \citep[][]{di_121102, k_121102, platts2021, Marazuela2020, Law2017}. The advantage of assuming a functional form for the spectra is that it enables the estimation of fluence and bandwidth of the burst signal, even if the whole burst is not visible in the band \citep[][]{k_121102}. This can provide a robust estimate of the intrinsic fluence and bandwidth of the burst if the assumed functional form is correct.    

\subsection{Estimating intrinsic properties}
It is now possible to establish a hierarchical framework to infer the intrinsic properties of the repeater bursts. Such a framework would require the following ingredients: 1) observed (or preferably fitted) fluences of the detected bursts from a repeater, 2) completeness limit of the observing system estimated using rigorous injection analysis \citep{gb, gupta2021}, 3) DM grid (or DM tolerance) used in single pulse search, 4) Boxcar widths searched. It would need to assume a distribution for burst energies, a spectral shape, and optionally an intrinsic DM and width distribution.


\subsection{Effects of power-law slope}
We also tested the effect of power-law slope on the energy and spectral parameter distributions of the detected bursts. We re-analyzed the example listed above with two more power-law slopes: $-1.2$ and $-1.8$, and observed the recovered distributions. We did not detect any significant difference in the results for these two power-law slopes with respect to the ones presented earlier, for a slope of $-1.5$. The cumulative distribution of energy showed a similar flattening and turnover with decreasing sensitivity. 
The in-band bursts still provided a more robust and accurate estimate of the intrinsic slope. The burst energy distributions also showed similar shapes as reported in Section~\ref{sec:energy_pdf}. 

\section{Conclusions}
In this Letter, we have discussed various observational effects that can arise due to the banded nature of the spectra of repeating FRBs. Primarily, the banded nature of burst spectra leads to a non-uniform completeness limit across the observing band. This is because many bursts that lie primarily outside the observing band will not be detected. Therefore, contrary to what is typically understood, the search pipeline is not complete to all the fluences above the sensitivity limit determined using traditional injection analysis \citep{gb, gupta2021, Marazuela2020}. This incompleteness must be accounted in the analysis that is used to determine the intrinsic properties of the FRB. Also, it is challenging to estimate the intrinsic fluence and bandwidth of bursts that lie on the edge of the observing band. We assumed the spectra to have a Gaussian shape and simulated bursts from an FRB~121102-like source. We then showed that the energy distribution of the detected bursts looks substantially different from the intrinsic distribution and might show peculiar shapes. We also showed that modeling the burst spectra using a Gaussian to determine the intrinsic fluence and bandwidth provides more robust results than traditional approaches. 

A normal intrinsic distribution of $\mu_f$ and $\sigma_f$ can explain the observed distribution seen for FRB~121102 \citep[][]{k_121102}, if the peak of the $\mu_f$ distribution lies towards the higher frequency end of the observing band. We point out that burst bandwidths, instead of center frequency, should be used to estimate the energy of the banded repeaters from fluence. We also showed that the bimodality in the energy distribution of FRB~121102 bursts reported by \citet[][]{di_121102} disappears when the energy is estimated using burst bandwidths instead of the center frequency of the band. Based on our tests, we can make the following recommendations for single pulse search and analysis of repeater bursts:
\begin{enumerate}
    \item Fluence and energies derived using fitting should be preferred over the ones estimated from signal to noise of the burst. Moreover, only the bursts whose peak and bulk of the emission lie primarily within the observing band (i.e., the in-band bursts) should be used to make inferences about the intrinsic distribution of energies and other properties of the FRB.   
    \item If the cumulative energy distribution shows a break in the power-law, then the higher energy power-law could follow the intrinsic distribution. A smooth turnover in the power-law will probably not represent the intrinsic distribution in the absence of a break. Even still, if the observations are not very sensitive, it might be impossible to recover the intrinsic properties of the repeater (see Section~\ref{sec:cumulative_e}). Moreover, these effects depend on the observed fluences from the source. An energetic repeater, which is also close to us, might be easier to interpret than one that is further away.
    \item Subbanded searches are more sensitive to such band-limited bursts and will aid in resolving some of the observational biases listed in this Letter.
    \item Analysis to determine the search pipeline completeness should incorporate band-limited spectra of FRBs in the simulated FRB injections.
\end{enumerate}
We note that these conclusions apply only to band-limited transient emission, i.\,e. they might not necessarily apply to situations where the intrinsic burst bandwidth is much greater than the bandwidth of the observing hardware in use.
All the analysis scripts and notebooks used in this work are provided in a Github repository\footnote{\url{https://github.com/KshitijAggarwal/banded_repeater_analysis}}.

\begin{acknowledgments}
We would like to thank Devansh Agarwal, Sarah Burke-Spolaor and Casey J. Law for useful discussions and comments on the manuscript. We also thank Di Li for sharing the properties of FRB~121102 bursts detected with FAST. K.A. acknowledges support from NSF grants AAG-1714897 and \#2108673. 
\end{acknowledgments}

%

\vspace{5mm}


\software{Astropy \citep{astropy:2013, astropy:2018}, Numpy \citep{harris2020}, Matplotlib \citep{Hunter:2007}, Pandas \citep{pandas2010, reback2020pandas}, SciPy \citep[][]{2020SciPy-NMeth}, SciencePlots \citep{SciencePlots}}



\bibliography{observational}{}
\bibliographystyle{aasjournal}



\end{document}